\documentclass{aa}  

\usepackage{graphicx, color}
\usepackage{txfonts}
\usepackage{comment}
\usepackage{siunitx}  
\usepackage{booktabs} 
\usepackage{orcidlink}
\begin{document}

   \title{Spectropolarimetric characterisation of exoplanet host stars in preparation of the $Ariel$ mission}
\subtitle{Magnetic environment of HD\,63433}
   \titlerunning{The magnetic environment of HD\,63433}

   \author{S. Bellotti \inst{1,2}\orcidlink{0000-0002-2558-6920}
          \and
          D. Evensberget \inst{1}\orcidlink{0000-0001-7810-8028}
          \and
          A. A. Vidotto \inst{1}\orcidlink{0000-0001-5371-2675}
          \and
          A. Lavail \inst{2}\orcidlink{0000-0001-8477-5265}
          \and
          T. L\"uftinger \inst{3}\orcidlink{0009-0000-4946-6942}
          \and
          G. A. J. Hussain \inst{3}\orcidlink{0000-0003-3547-3783}
          \and
          J. Morin \inst{4}\orcidlink{0000-0002-4996-6901}
          \and
          P. Petit \inst{2}\orcidlink{0000-0001-7624-9222}
          \and
          S. Boro Saikia \inst{5}\orcidlink{0000-0002-3673-3746}
          \and
          C. Danielski \inst{6}\orcidlink{0000-0002-3729-2663}
          \and
          G. Micela \inst{7}\orcidlink{0000-0002-9900-4751}
          }
   \authorrunning{Bellotti et al.}
    
   \institute{
            Leiden Observatory, Leiden University,
            PO Box 9513, 2300 RA Leiden, The Netherlands\\
            \email{bellotti@strw.leidenuniv.nl}
        \and
            Institut de Recherche en Astrophysique et Plan\'etologie,
            Universit\'e de Toulouse, CNRS, IRAP/UMR 5277,
            14 avenue Edouard Belin, F-31400, Toulouse, France  
        \and
             Science Division, Directorate of Science, 
             European Space Research and Technology Centre (ESA/ESTEC),
             Keplerlaan 1, 2201 AZ, Noordwijk, The Netherlands
        \and
             Laboratoire Univers et Particules de Montpellier,
             Universit\'e de Montpellier, CNRS,
             F-34095, Montpellier, France
        \and University of Vienna, Department of Astrophysics, Türkenschanzstrasse 17, A-1180 Vienna, Austria
        \and INAF - Osservatorio Astrofisico di Arcetri, Largo E. Fermi 5, 50125, Firenze, Italy
        \and INAF - Osservatorio Astronomico di Palermo, Piazza del Parlamento 1, 90134, Palermo, Italy
             }
   \date{Received ; accepted }

 
  \abstract
   {The accurate characterisation of the stellar magnetism of planetary host stars has been gaining momentum, especially in the context of transmission spectroscopy investigations of exoplanets. Indeed, the magnetic field regulates the amount of energetic radiation and stellar wind impinging on planets, as well as the presence of inhomogeneities on the stellar surface that hinder the precise extraction of the planetary atmospheric absorption signal.}   
   {We initiated a spectropolarimetric campaign to unveil the magnetic field properties of known exoplanet hosting stars included in the current list of potential $Ariel$ targets. In this work, we focus on HD\,63433, a young solar-like star hosting two sub-Neptunes and an Earth-sized planet. These exoplanets orbit within 0.15\,au from the host star and have likely experienced different atmospheric evolutionary paths. }
   {We analysed optical spectropolarimetric data collected with ESPaDOnS, HARPSpol, and Neo-Narval to compute the magnetic activity indices ($\log R'_\mathrm{HK}$, H$\alpha$, and Ca~\textsc{ii} infrared triplet), measure the longitudinal magnetic field, and reconstruct the large-scale magnetic topology via Zeeman-Doppler imaging (ZDI). The magnetic field map was then employed to simulate the space environment in which the exoplanets orbit.}
   {The reconstructed stellar magnetic field has an average strength of 24\,G and it features a complex topology with a dominant toroidal component, in agreement with other stars of a similar spectral type and age. Our simulations of the stellar environment locate 10\% of the innermost planetary orbit inside the Alfvén surface and, thus, brief magnetic connections between the planet and the star can occur. The outer planets are outside the Alfvén surface and a bow shock between the stellar wind and the planetary magnetosphere could potentially form.}
   {}

   \keywords{Stars: magnetic field --
                Stars: individual: HD\,63433 --
                Stars: activity --
                Techniques: polarimetric
               }

   \maketitle

%

\section{Introduction}

The field of exoplanetology has flourished significantly over the past three decades. With more than $5,500$ confirmed detections\footnote{\url{www.exoplanet.eu}, December 2023}, the main goal is to progressively focus on a comprehensive characterisation of the planetary population. Our knowledge of planet formation and evolution will indeed progress by inspecting the chemical variety of exoplanet atmospheres and, in turn, it will allow us to constrain planet formation theories and refine the assessment framework of potential biomarkers. These considerations motivated the deployment of space-based missions currently in operation such as $JWST$ \citep{Gardner2006}, as well as future ones such as $Ariel$ \citep{Tinetti2021}, which is a medium-class science mission planned for launch by the European Space Agency in  2029. It will target a comprehensive and diverse sample of telluric, ocean, and gas giant planets discovered primarily via radial velocities and transit photometry, orbiting around stars of all accessible spectral types, from early A to late M \citep{Zingales2018,Edwards2019,Edwards2022} types.

When it comes to atmospheric characterisation surveys of exoplanets, stellar magnetic activity is a crucial aspect that ought to be taken into account to avoid wrongful or biased interpretations of the planetary nature. In fact, stellar activity is responsible for photospheric inhomogeneities such as spots and faculae, to which the optical and near-infrared (NIR) are sensitive within the spectral ranges commonly examined for exoplanet detection and characterisation.  These inhomogeneities contaminate radial velocity signals \citep[e.g.][]{Huelamo2008,Dumusque2011}, either mimicking a planetary signal \citep[e.g.][]{Carmona2023} or preventing an accurate estimation of the planetary mass, ultimately affecting a correct atmospheric retrieval \citep{Changeat2020,DiMaio2023}. In the case of transmission spectroscopy, stellar activity can bias the observed spectrum, causing the transit chord to differ from the disk-integrated stellar spectrum, generating ambiguity in the extraction of the planetary atmospheric absorption signal \citep[e.g.][]{Rackham2018,Rackham2019,Salz2018}. For instance, spots affect the transit depth with at least the same order of magnitude as transiting exoplanets, potentially emulating the presence of specific chemical species like H$_2$O \citep{Rackham2018}.
Therefore, designing activity-filtering techniques for transmission spectroscopy is necessary \citep{Cracchiolo2021,Thompson2023}.

In addition, magnetic activity shapes the environment in which exoplanets orbit, and it affects their atmospheres, especially during the early stages of their formation \citep[e.g.][]{Ribas2005,Allan2019,Ketzer2023}. Strong activity can endanger habitability \citep[e.g.,][]{Airapetian2017,Tilley2019}, as it indeed correlates with energetic phenomena, such as flares, ultimately modifying the atmospheric chemical composition \citep{Segura2010,Guenther2020,Konings2022,Louca2022}. Ionising radiation (e.g. in the extreme-ultraviolet and X-rays) and particles from strong stellar winds induce and shape hydrodynamic escape of the atmosphere, stripping it away over time \citep{Lammer2003,Ribas2016,Carolan2021,Hazra2020}. XUV radiation also impacts secondary atmospheres, such as the case of Trappist-1 \citep{vanLooveren2024}. Moreover, \citet{Chavez2023} show that the cloud coverage on Neptune is temporally correlated with the solar magnetic cycle. For these reasons, there is increased attention towards a comprehensive and homogeneous characterisation of exoplanet hosts \citep[e.g.][]{Danielski2022,Magrini2022} and their magnetic activity \citep{Fares2013,Mengel2016,Fares2017,Folsom2020}. A recent example is given by our work on GJ~436 and its orbiting hot Neptune \citep{Bellotti2023a,Vidotto2023}.

To obtain a complete picture of the magnetic activity of stars in the current list of potential $Ariel$ targets \citep{Edwards2019,Edwards2022}, we initiated a dedicated spectropolarimetric programme divided into three steps: 1) a snapshot campaign to assess the detectability of the large-scale magnetic field and optimise further observations; 2) an observing campaign to reconstruct the topology of the surface stellar magnetic field; and 3) a long-term monitoring to constrain the evolution and variability of the field. Spectropolarimetry enables us to achieve these goals effectively, as we detect the polarisation properties of the Zeeman-split components of spectral lines \citep{Zeeman1897}. From the time series of polarised spectra, we mapped the large-scale magnetic field with Zeeman-Doppler imaging (ZDI; \citealt{Semel1989,Donati1997}), constraining its configuration and main features. This technique has been applied extensively in spectropolarimetric studies, revealing a variety of field geometries for cool stars \citep[e.g.][]{Petit2005,Hussain2016,Morin2008,Morin2010} and long-term evolution \citep{BoroSaikia2016,BoroSaikia2018,Bellotti2023b}.

This paper focuses on HD\,63433, a young, active solar-like star which is part of the current list of potential $Ariel$ targets and hosts two transiting mini Neptunes (planets b and c orbiting at 0.07 and 0.15\,AU; \citealt{Mann2020,Mallorquin2023,Damasso2023}). As reported by transmission spectroscopy studies of their atmospheres \citep{Zhang2022}, both planets do not exhibit helium absorption, whereas only planet c shows Ly$\alpha$ absorption. Considering the architecture of the system, along with the age and activity of the star, this suggests that HD\,63433\,b has probably lost most of its primordial H/He atmosphere, while HD\,63433\,c is still experiencing atmospheric evaporation. These results make the HD\,63433 system an attractive benchmark for comparative studies of distinct atmospheric evolution tracks and their timescales. Recently, a third planet with an orbital period of 4.2\,d was announced \citep{Capistrant2024} and its proximity to the star makes it a potentially interesting target for star-planet interaction studies.

The paper is structured as follows. In Sect.~\ref{sec:observations}, we describe the observations obtained with ESPaDOnS and HARPS-Pol, and the subsequent Neo-Narval monitoring. In Sect.~\ref{sec:mag_characterisation}, we outline the computation of the activity proxies, longitudinal magnetic field, and their correlation analysis. We present in Sect.~\ref{sec:zdi} the large-scale magnetic field reconstruction by means of ZDI, and in Sect.~\ref{sec:wind} the ZDI-driven stellar wind models. Finally, we draw our conclusions in Sect.~\ref{sec:discussion}.

\section{Observations}\label{sec:observations}

HD\,63433 is a G5\,dwarf at a distance of 22.34$\pm$0.04\,pc \citep{Gaia2020}. It is a bright ($V=6.9$) star belonging to the Ursa Major moving group, with an estimated age of $414\pm23$\,Myr \citep{Gagne2018,Mann2020}. The star has a rotation period of 6.45\,d, and is magnetically active, with a reported chromospheric activity index $\log R'_\mathrm{HK}=-4.39\pm0.05$ and an X-ray-to-bolometric luminosity index $\log(L_X/L_\mathrm{bol})=-4.04\pm0.13$ \citep{Mann2020}. The $L_X/L_\mathrm{bol}$ is two orders of magnitude larger than the solar value \citep{Wright2011}, while the $\log R'_\mathrm{HK}$ is 0.5\,dex larger \citep{Egeland2017}. A summary of the fundamental parameters of HD\,63433 is given in Table~\ref{tab:hd63433_properties}. Two sub-Neptunes were discovered to orbit the star via photometric transits \citep{Mann2020}: HD\,63433\,b with a radius and orbital period of 2.02\,R$_\oplus$ and 7.11\,d, and HD\,63433\,c, with a radius of 2.44\,R$_\oplus$ and on a 20.55\,d orbit. Radial velocity follow-up of this system by \citet{Damasso2023} resulted in planetary mass upper limits of 11\,M$_\oplus$ for planet b and 31\,M$_\oplus$ for planet c. The work of \citet{Mallorquin2023} set a mass of 15.5\,M$_\oplus$ for planet c and an upper limit of 22\,M$_\oplus$ for planet b. A third planet was also reported recently via photometric transit, with a radius of 1.073\,R$_\oplus$ and orbital period of 4.2\,d \citep{Capistrant2024}. 

\subsection{ESPaDOnS}

We obtained one observation with ESPaDOnS\footnote{\href{https://www.cfht.hawaii.edu/Instruments/Spectroscopy/Espadons/}{https://www.cfht.hawaii.edu/Instruments/Spectroscopy/Espadons/}} on 12 September 2022, within the snapshot campaign (22BF23, P.I. Bellotti S.) of the spectropolarimetric characterisation programme (see Table~\ref{tab:log}). ESPaDOnS is the optical spectropolarimeter on the 3.6~m CFHT located atop Mauna Kea in Hawaii, operating between 370 and 1050\,nm with a spectral resolving power of $R=65,000$ \citep{Donati2003}. A polarimetric sequence is obtained from four consecutive sub-exposures. Each sub-exposure is taken with a different rotation of the retarder waveplate of the polarimeter relative to the optical axis. In circular polarisation mode, the output is given by the total intensity spectrum (Stokes~$I$) and the circularly polarised spectrum (Stokes~$V$). The Stokes~$I$ spectrum is computed by summing the four sub-exposures, while the Stokes~$V$ spectrum from the ratio of sub-exposures with orthogonal polarisation states. Circular polarisation mode is more appropriate for snapshot campaigns of main sequence cool dwarf stars, since the amplitude of a Zeeman signature is larger than in linear polarisation \citep{Landi1992}. The data reduction was performed with the \textsc{LIBRE-ESPRIT} pipeline \citep{Donati1997}, and the maximum signal-to-noise ratio (S/N) per CCD pixel of the circularly polarised spectrum is 735.

We applied least-squares deconvolution (LSD) \citep{Donati1997,Kochukhov2010a} to compute the Stokes~$I$ and $V$ mean-line profiles. Both the unpolarised and polarised spectra are deconvolved with a line mask, which is a list of photospheric absorption lines for the specific spectral type examined. Least-squares deconvolution results in high-S/N line profiles encapsulating the average information of thousands of lines. We used a line mask synthesised using the Vienna Atomic Line Database\footnote{\url{http://vald.astro.uu.se/}} \citep[VALD,][]{Ryabchikova2015}, characterised by $\log g=$ 4.5\,[cm s$^{-2}$], $v_{\mathrm{micro}}=$ 1\,km s$^{-1}$, and $T_{\mathrm{eff}}=5500$\,K. The mask includes 4620 atomic lines between 360--1080\,nm, with known sensitivity to the Zeeman effect (also known as Land\'e factor and indicated as $g_\mathrm{eff}$) and an absorption depth greater than 40\,\% of the continuum level. This value is chosen to obtain a larger effective S/N of the LSD profiles \citep{Moutou2007}. The normalisation wavelength and the Land\'e factor for LSD are 700\,nm and 1.0, respectively.

\begin{figure}
    \includegraphics[width=\columnwidth]{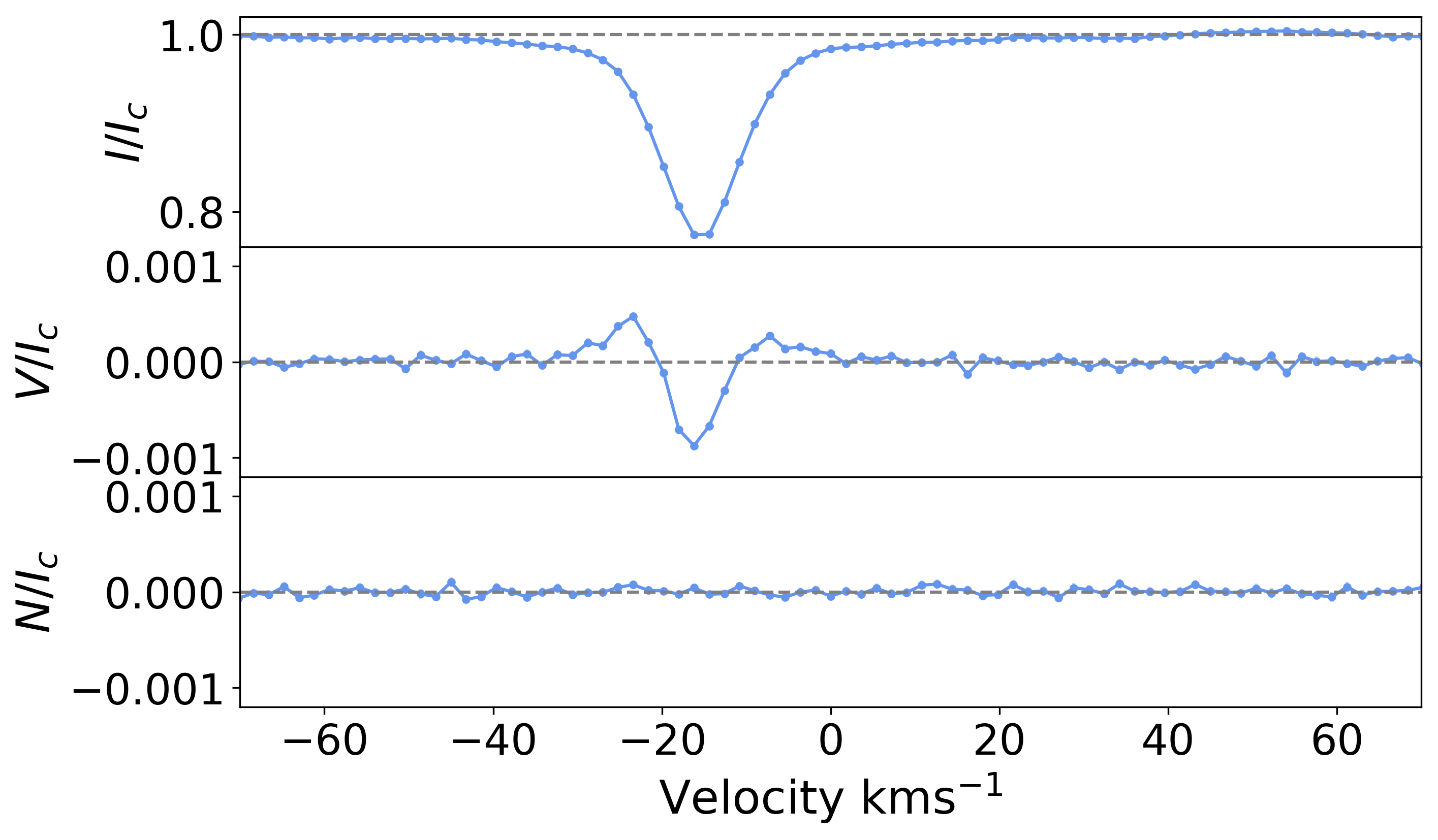}
    \caption{Least-squares deconvolution profiles for the snapshot observations of HD\,63433 with ESPaDOnS. The panels show the Stokes~$I$ (top), $V$ (middle) and $N$ (bottom) profiles, obtained from the combination of 4620 atomic spectral lines. We note the presence of a clear Zeeman signature in Stokes~$V$, indicating that the large-scale magnetic field is amenable to characterisation.
        }\label{fig:stokes_snap}
\end{figure}

The Stokes LSD profiles for the snapshot observation are shown in Fig.~\ref{fig:stokes_snap}. Owing to an LSD multiplex gain of about 30, the S/N per 1.8\,km\,s$^{-1}$ velocity bin of the Stokes~$V$ profile is $\sim$20,000. We note an evident Zeeman signature in Stokes~$V$ within $\pm$17\,km\,s$^{-1}$ from line centre at $-15.9$~km\,s$^{-1}$, and with an associated false-alarm probability lower than $10^{-5}$, indicating a reliable magnetic detection \citep{Donati1997}. The polarimetric pipeline provides the null spectrum (Stokes~$N$) as well, obtained by dividing sub-exposures with the same polarisation state. We computed the associated LSD profile, with which we can assess the presence of spurious polarisation signatures and the overall noise level in the LSD output \citep{Donati1997,Bagnulo2009}. Within $\pm$17\,km\,s$^{-1}$ from line centre, the Stokes~$N$ profile has a mean of \qty{4.7e-6}, a standard deviation of \qty{3.6e-5}, and a false-alarm probability of \qty{7.9e-1}, confirming the absence of spurious polarimetric signatures. The success of this preliminary spectropolarimetric snapshot of HD\,63433 represents the gateway for a subsequent monitoring dedicated to the characterisation of the large-scale magnetic field.

Our observations were carried out in circular polarisation mode only, considering the prohibitive exposure time required for linear polarisation mode for main sequence stars such as HD\,63433. The inclusion of linear polarisation would allow us to reduce the cross-talk between the radial and meridional field components and obtain a better constrain of smaller magnetic field features \citep[e.g.][]{Rosen2015}.

\begin{table}[!t]
\caption{Properties of HD\,63433 that are relevant for our study.} 
\label{tab:hd63433_properties}     
\centering                       
\begin{tabular}{l c c}    
\toprule
Parameter & Value & Reference\\
\midrule
B [mag] & $7.59\pm0.02$ & [1]\\
V [mag] & $6.91\pm0.01$ & [1]\\
Distance [pc] & 22.34$\pm$0.04 & [2]\\
T$_\mathrm{eff}$ [K] & $5700\pm75$ & [3]\\
$\log g$ [dex] & $4.54\pm0.05$ & [3]\\
Mass [M$_\odot$] & $0.99\pm0.03$ & [3]\\
Radius [R$_\odot$] & $0.897\pm0.019$ & [3]\\
Age [Myr] & $414\pm23$ & [4]\\
P$_\mathrm{rot}$ [d] & $6.45\pm0.05$ & [4]\\
$v_\mathrm{eq}\sin i$ [km\,s$^{-1}$] & $7.3\pm0.3$ & [4]\\
Inclination [$^\circ$] & $>74$ & [4]\\
$S$-index & $0.433\pm0.041$ & [5]\\
$\log R'_\mathrm{HK}$ [dex] & $-4.345\pm0.055$ & [5]\\
\bottomrule 
\end{tabular}
\tablefoot{References: (1) \citet{Hog2000}, (2) \citet{Gaia2020}, (3) \citet{Damasso2023}, (4) \citet{Mann2020}, (5) This work.}
\end{table}

\begin{table*}[!t]
\caption{Our observations of HD\,63433.} 
\label{tab:log}     
\centering                       
\begin{tabular}{c c r c c c c r}    
\toprule
Date & UT & HJD & $n_\mathrm{cyc}$ & Instrument & $t_\mathrm{exp}$ & S/N & $\sigma_\mathrm{LSD}$ \\
{[dd-mm-yyyy]} & [hh:mm:ss] & [$-2450000$] & & & [s] & & [$10^{-5}I_c$]\\
\midrule
12-09-2022 & 15:01:01.380 & 9835.12079 & $-$27.17 & ESPaDOnS & 4x340 & 743 & 4.83 \\
23-12-2022 & 05:32:01.248 & 9936.73575 & $-$11.42 & HARPSpol & 4x900 & 394 & 13.70 \\
25-12-2022 & 05:23:12.480 & 9938.72971 & $-$11.11 & HARPSpol & 4x900 & 327 & 201.53 \\
06-02-2023 & 04:22:00.480 & 9981.68718 & $-$4.45 & HARPSpol & 4x900 & 245 & 11.42 \\
06-03-2023 & 20:55:29.485 & 10010.37541 & 0.00 & Neo-Narval & 4x900 & 520 & 9.60\\
07-04-2023 & 19:46:09.397 & 10042.32441 & 4.95 & Neo-Narval & 4x900 & 556 & 11.70\\
08-04-2023 & 20:17:03.480 & 10043.34576 & 5.11 & Neo-Narval & 4x900 & 465 & 8.71\\
16-04-2023 & 19:25:34.170 & 10051.30923 & 6.35 & Neo-Narval & 4x900 & 494 & 9.83\\
02-05-2023 & 19:49:12.432 & 10067.32409 & 8.83 & Neo-Narval & 4x900 & 381 & 13.60\\
05-05-2023 & 20:45:55.940 & 10070.36319 & 9.30 & Neo-Narval & 4x900 & 417 & 9.44\\
\bottomrule
\end{tabular}
\tablefoot{The following quantities are listed: 1-2) date and time of the observation; 3) Heliocentric Julian date; 4) cycle number based on the ephemeris in Eq.~\ref{eq:ephemeris}; 5) instrument used; 6) exposure time per polarimetric sequence; 7) S/N per polarimetric sequence at 650\,nm per CCD pixel of unpolarised spectrum; and 8) RMS noise level of Stokes $V$ signal in units of unpolarised continuum.}
\end{table*}

\subsection{HARPSpol}

HD~63433 was observed with HARPSpol\footnote{\href{https://www.eso.org/sci/facilities/lasilla/instruments/harps.html}{https://www.eso.org/sci/facilities/lasilla/instruments/harps.html}} \citep{2011ASPC..437..237S, 2011Msngr.143....7P}, the spectropolarimeter for the HARPS spectrograph \citep{2003Msngr.114...20M} at the ESO 3.6~m telescope at La Silla observatory, Chile. The HARPSpol instrument is mounted at the Cassegrain focus of the telescope and has two polarimeters:\ for circular (Stokes $V$) and linear (Stokes $QU$) polarisation, respectively. Each polarimeter consists of a rotating retarder wave plate (a quarter-wave plate for the circular polarimeter and a half-wave plate for the linear polarimeter) and a beam-splitter -- in this case a Foster prism -- which splits the light into two beams with opposite polarisation states. These two beams are then fed via fibers to the HARPS spectrograph. Our HARPSpol observations cover the wavelength range between 380 and and 691~nm with a 8~nm gap at 529~nm separating the red and blue detectors. The spectral resolution is $ R = \delta \lambda / \lambda \approx 110 000$. We obtained circular polarisation observations on three separate nights: 22 December 2022, 24 December 2022, and 06 February 2023.

To reduce the HARPSpol data, we used the {\tt PyReduce} package\footnote{\url{https://github.com/AWehrhahn/PyReduce}} \citep{2021A&A...646A..32P}, the updated implementation in {\tt python} of the versatile {\tt REDUCE} package \citep{2002A&A...385.1095P}. The reduction with {\tt PyReduce} is run with a series of standard steps, similar to {\tt REDUCE} as described, for instance, \citet{2013A&A...558A...8R}. First, the mean of both bias and flat frames is computed, followed by the substraction of the mean bias from the mean flat. The final mean flat frame is then used to compute the normalized flat field, and to determine the location and shape of the spectral orders on the detectors. Science frames first get the bias subtracted and are divided by the normalized flat field to correct pixel-to-pixel variation in sensitivity. Scattered light is then subtracted from the science frames. Finally, the spectra from the two polarisation channels are extracted for each science frame. At this point, the extracted science spectra are in pixel space. To convert from pixel to wavelength space, we constructed a 2D wavelength solution from the wavelength calibration frames taken with the Thorium-Argon (ThAr) lamp as part of the regular daily calibrations. In this reduction, the blue and red detectors of HARPS are reduced independently.

We then applied LSD to the collected HARPSpol spectra using the same mask as ESPaDOnS. Compared to ESPaDOnS, the final LSD profiles are expected to be penalised due to the narrower spectral coverage of HARPSpol. For the night of 22 December 2022, we detected a Zeeman signature with FAP \qty{8.7e-9}, and the Stokes~$V$ S/N is 7900. The two other nights do not show detections (FAP is \qty{9.4e-1} and \qty{3.8e-2}) and they correspond to the observations with the lowest S/N (see Table~\ref{tab:log}). Therefore, while we computed the longitudinal field only for the first observation, we measured the chromospheric activity indices for all three observations. Given that the spectral coverage of HARPSpol stops at 691~nm, we did not compute the Ca~\textsc{ii} infrared triplet activity diagnostics (see Sect.~\ref{sec:mag_characterisation}).

\subsection{Neo-Narval}

HD\,63433 was observed by Neo-Narval\footnote{\href{https://tbl.omp.eu/instruments/neo-narval/}{https://tbl.omp.eu/instruments/neo-narval/}} between March and May 2023. We obtained six observations and although the phase coverage is not ideal, it is sufficient to reconstruct the topology of the large-scale magnetic field. The complete time series is summarised in Table~\ref{tab:log}. Neo-Narval is an upgrade to the Narval instrument at T\'elescope Bernard Lyot (Pic du Midi Observatory, \citealt{Auriere2003}), operating between 370 and 1000\,nm with a spectral resolving power of $R=65,000$.

The Neo-Narval upgrade, implemented in 2019, included the installation of a new detector and improved velocimetric capabilities \citep{LopezAriste2022}. Issues with the extraction of blue spectral orders occurred during the period of our observations \citep{LopezAriste2022}, affecting especially those with very low S/N (below approximately 400\,nm). As explained in the next section, this prevented us from studying \textsc{Ca II H\&K} lines.
In contrast, the computation of LSD profiles was less affected because LSD uses a S/N-weighting scheme. We adopted the same synthetic mask of 4620 lines as for ESPaDOnS and normalisation parameters, obtaining Stokes~$V$ profiles with the maximum S/N spanning between 7300 and 11,480. In the next sections, all the observations will be phased with the following ephemeris
\begin{equation}
    \mathrm{JD} = 2460010.37186 + \mathrm{P}_\mathrm{rot}\cdot n_\mathrm{cyc},\\
    \label{eq:ephemeris}
\end{equation}
where we used the first Neo-Narval observation as JD reference, P$_\mathrm{rot}$ is the stellar rotation period, and $n_\mathrm{cyc}$ represents the rotation cycle (see Table~\ref{tab:log}).

\section{Magnetic activity proxies}\label{sec:mag_characterisation}

\subsection{Activity indices}

The spectral coverage of both ESPaDOnS, HARPSpol, and Neo-Narval encompasses chromospheric lines that are typically used to gauge chromospheric flux and thus assess stellar activity. As described in the next sections, we computed canonical activity indices using the Ca~\textsc{ii} H\&K lines, the H$\alpha$ line and Ca~\textsc{ii} infrared triplet lines. In the absence of star-planet interactions, these indices should vary at the stellar rotation period timescale \citep[e.g.][]{Mittag2017,Kumar2023}, although it is known that chromospheric complexity hampers rotation period measurements. For this reason, and considering the low number of observations, we did not perform a temporal analysis.

\subsubsection{Ca II H\&K index}

Stellar activity has been quantified with the $S$~index \citep{Vaughan1978} and it was firstly adopted to search for activity cycles within the Mount Wilson project \citep{Wilson1968,Duncan1991}. It is defined as the ratio between the flux of the Ca~\textsc{ii} H\&K lines and the flux of the nearby continuum. Formally, 
\begin{equation}
    S = \frac{aF_H+bF_K}{cF_R+dF_V}+e,
\end{equation}
where $F_H$ and $F_K$ are the fluxes in two triangular band passes with FWHM = 1.09\,{\AA} centred on the cores of the H line (3968.470\,{\AA}) and K line (3933.661\,{\AA}), whereas $F_R$ and $F_V$ are the fluxes within two 20-{\AA} rectangular band passes centred at 3901 and 4001\,{\AA}, respectively. The set of coefficients $\{a,b,c,d,e\}$ is used to convert the $S$-index from a specific instrument scale to the Mount Wilson scale. The coefficients for the ESPaDOnS instrument were estimated by \citet{Marsden2014}, while for HARPSpol we used the optimisation of \citet{BoroSaikia2018}. Given the Neo-Narval pipeline issue for the spectral region below 400\,nm, we did not compute the activity index for the corresponding observations.

We determined the $S$-index from the intensity spectra of all four sub-exposures composing the ESPaDOnS snapshot observation (see Sect.~\ref{sec:observations}), which we Doppler-shifted according to the radial velocity of the star ($-15.9$\,km\,s$^{-1}$). We found a mean value of $0.433\pm0.021$, where the error bar is obtained as standard deviation between the four sub-exposures divided by two (i.e. the square root of the number of sub-exposures). We took this approach since the measurement of an activity indicator for a particular night has the benefit of an increased S/N thanks to the combination of sub-exposures. The value of $S$-index remains consistent if computed directly from the intensity spectrum obtained as combination of the four sub-exposures. The S/N in the spectral region where the Ca~\textsc{ii} H\&K lines lie is too low in the HARPSpol observations for a reliable measurement.

The $S$-index contains a colour dependence that biases the comparison of activity levels between stars, as well as a photospheric contribution in the line wings due to magnetic heating \citep{Noyes1984}. For this reason, the chromospheric activity index $\log R'_\mathrm{HK}$ is typically used \citep{Middelkoop1982,Noyes1984,Rutten1984}. The conversion yielded a value of $-4.345\pm0.027$ for the ESPaDOnS observation.

The index values we estimated from the ESPaDOnS spectrum are consistent with the range for which large-scale magnetic fields of active cool dwarfs are detectable \citep{Marsden2014}. They are compatible within uncertainties with the range reported by \citet{Pace2013} and \citet{BoroSaikia2018}, but they are larger by 0.06 dex (3$\sigma$) than the individual measurement in 2010 computed by \citet{Marsden2014}.

\subsubsection{H$\alpha$ index}

We measured the flux of the H$\alpha$ line normalised by the nearby continuum following \citet{Gizis2002}. This line is formed in the upper chromosphere, thus providing insights into this particular region of the stellar atmosphere. Formally,
\begin{equation}\label{eq:halpha}
    \mathrm{H}\alpha = \frac{F_{\mathrm{H}\alpha}}{H_R+H_V},
\end{equation}
where $F_{\mathrm{H}\alpha}$ is the flux within a rectangular band pass of 3.60\,{\AA} centred on the H$\alpha$ line at 6562.85\,{\AA}, and $H_V$ and $H_R$ are the fluxes within two rectangular band passes of 2.2\,{\AA} centred on 6558.85\,{\AA}  and 6567.30\,{\AA}.

The temporal evolution of the values computed from Neo-Narval spectra is shown in Fig.~\ref{fig:proxies} and the values are reported in Table~\ref{tab:indexes}. From the ESPaDOnS snapshot observation, we obtained a value of $0.341\pm0.002$, from HARPSpol we measured $0.339\pm0.001$, $0.339\pm0.001$, and $0.331\pm0.002$, and from Neo-Narval, we estimated values between 0.336 and 0.342 with a mean of $0.339\pm0.002$. Even though all observations spanned eight months in total, the activity level, as quantified by the H$\alpha$ index, did not indicate a substantial evolution.

\begin{figure}[!t]
    \centering
    \includegraphics[width=\columnwidth]{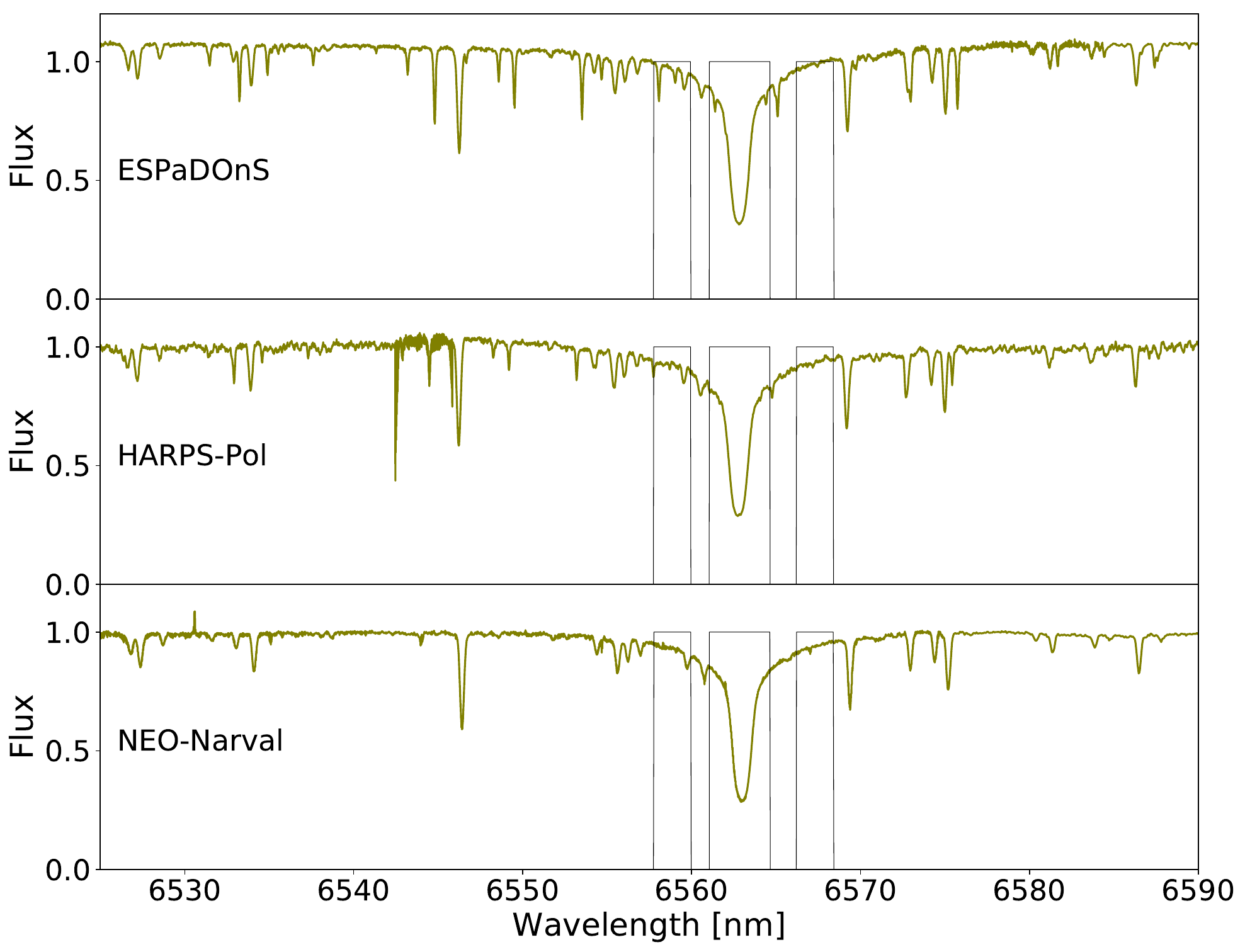}
    \caption{Normalised fluxes around the H$\alpha$ line for the three instruments used in this study. From the top: ESPaDOnS, HARPSpol, and Neo-Narval. The black rectangles indicate the regions used in the definition of the corresponding activity index. The observations were taken between September 2022 and May 2023.}
    \label{fig:halpha_mosaic}%
\end{figure}

\subsubsection{Ca II infrared triplet index}

Another chromospheric activity indicator is the Ca~\textsc{ii} infrared-triplet (IRT) index, which is a diagnostic for the lower chromosphere \citep{Montes2000,Hintz2019}. We followed \citet{Petit2013} and \citet{Marsden2014}, computing:
\begin{equation}\label{eq:irt}
    \mathrm{Ca II IRT} = \frac{\mathrm{IR1}+\mathrm{IR2}+\mathrm{IR3}}{\mathrm{IR}_R+\mathrm{IR}_V},
\end{equation}
where IR1, IR2, and IR3 are the fluxes within a rectangular band pass of 2\,{\AA} centred on the Ca~\textsc{ii} lines at 8498.023, 8542.091, and 8662.410\,{\AA}, respectively, and IR$_R$ and IR$_V$ are the fluxes within two rectangular band passes of 5\,{\AA} centred on 8704.9\,{\AA} and 8475.8\,{\AA}. The spectral coverage of HARPSpol does not encompass these calcium lines; hence, we did not compute the index for the corresponding observations.

\begin{figure}[!t]
    \centering
    \includegraphics[width=\columnwidth]{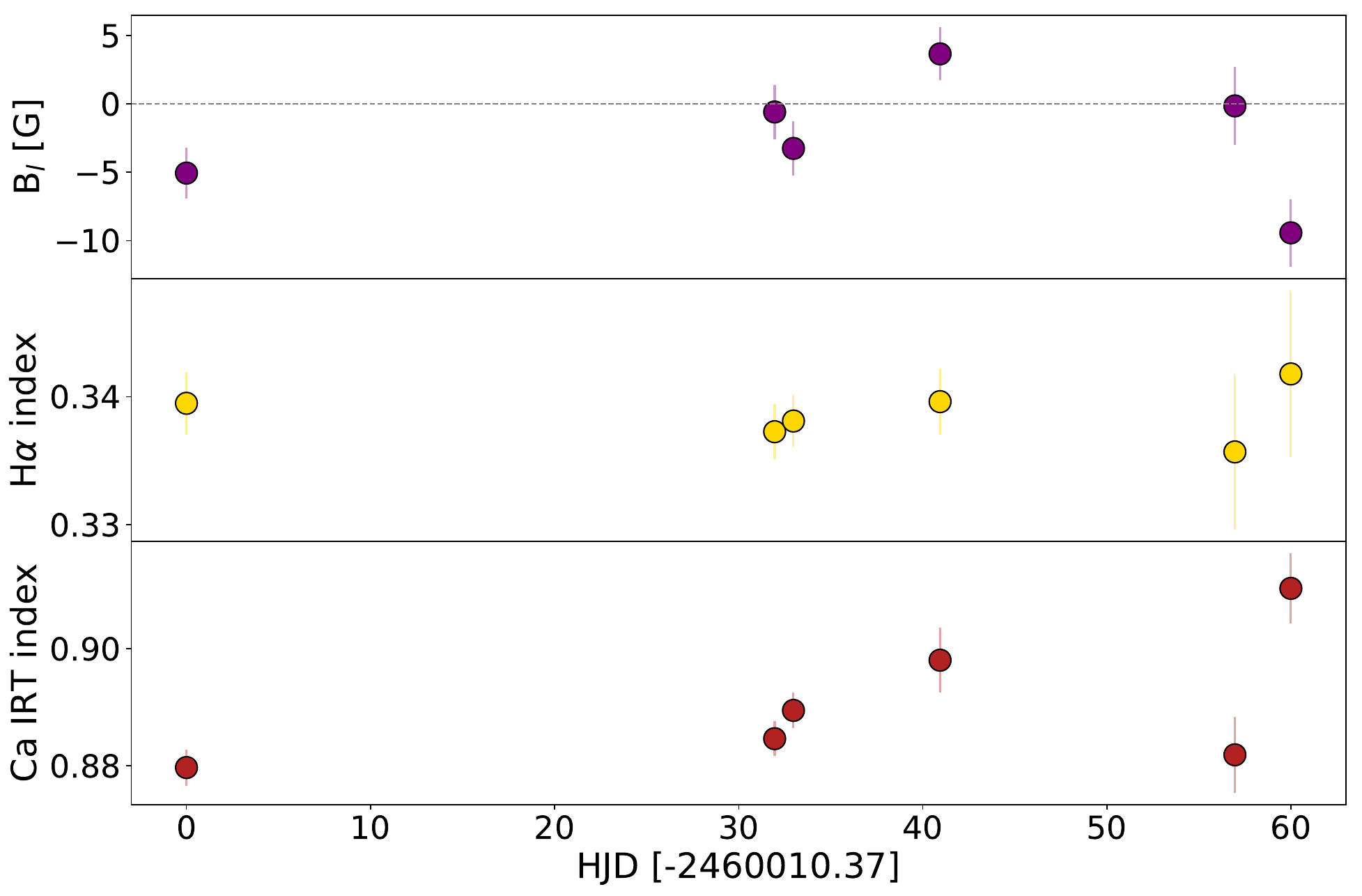}
    \caption{Time series of longitudinal magnetic field and activity indices measurements for HD\,63433 from the Neo-Narval observations. The longitudinal field (top panel) exhibits intermittency in the sign of the magnetic polarity, which is symbolic of a likely complex or non-axisymmetric topology. The H$\alpha$ index (middle panel) does not exhibit strong variations, while the Ca~\textsc{ii} IRT index (bottom panel) has an increased scatter towards the latest observations. The H$\alpha$ and Ca~\textsc{ii} IRT indices show correlated temporal evolution.}
    \label{fig:proxies}%
\end{figure}

The temporal evolution is shown in Fig.~\ref{fig:proxies} and the values reported in Table~\ref{tab:indexes}. From the ESPaDOnS snapshot observation, we obtained a value of $0.899\pm0.002$, whereas from the Neo-Narval observations, we estimated values between 0.880 and 0.910 with a mean of $0.891\pm0.002$. In a similar manner to H$\alpha$ index, the magnetic activity level seems to have remained reasonably stable, with the oscillations due to intrinsic variability of active regions. In general, we caution that robust conclusions on the evolution of the activity level based on the activity proxies computed here cannot be drawn, considering the large gaps between observations as well as the low number of observations.

\begin{figure*}[!t]
    \centering
    \includegraphics[width=\textwidth]{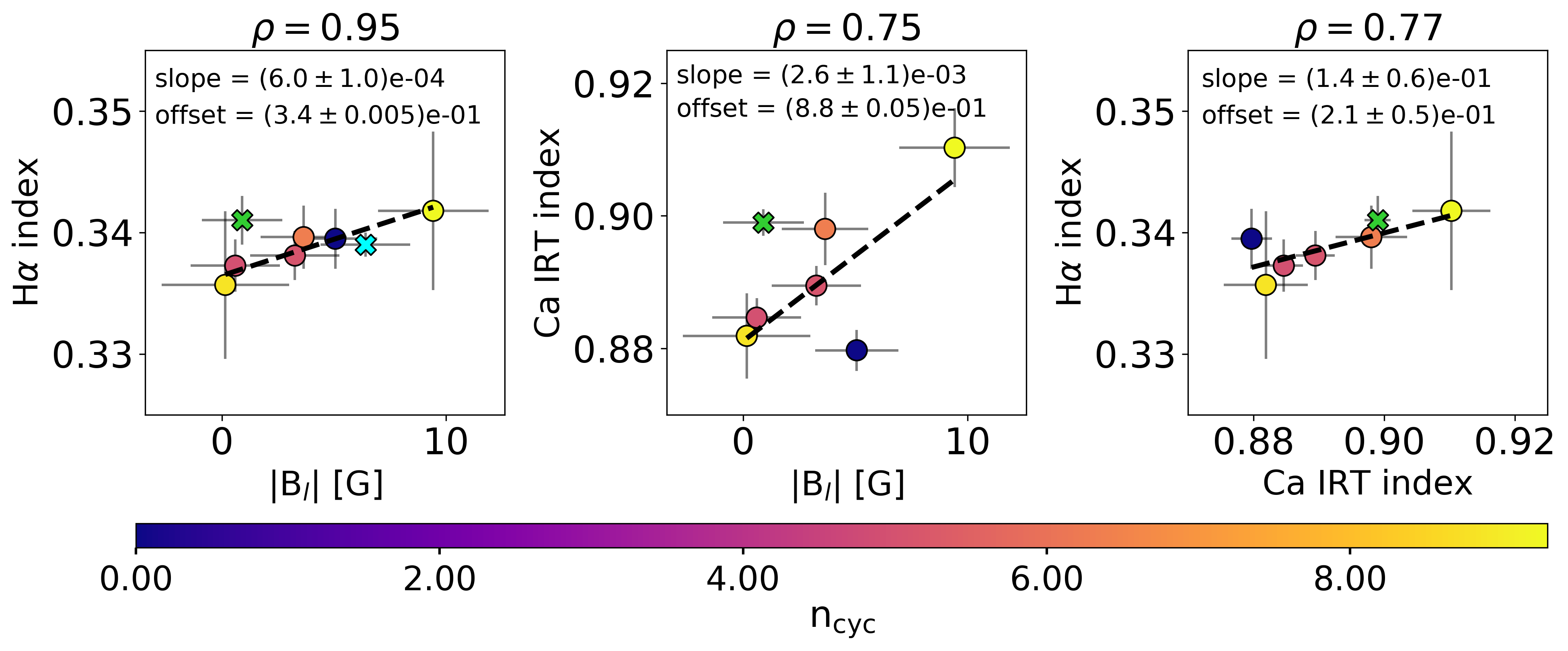}
    \caption{Correlation plots for the activity proxies computed from the six Neo-Narval observations. From the left: H$\alpha$ vs $|$B$_l|$, Ca~\textsc{ii} IRT vs $|$B$_l|$, and H$\alpha$ vs Ca~\textsc{ii} IRT. Each data point is colour-coded based on the rotational cycle as computed from Eq.\ref{eq:ephemeris}. Each panel displays the Pearson correlation coefficient ($\rho$) at the top and a linear fit as black dashed line. The green cross indicates the values of the ESPaDOnS snapshot observation, but it was not included in the computation of the Pearson correlation coefficient. We included the first HARPSpol observation in the first panel as a blue cross.}
    \label{fig:proxy_corr}%
\end{figure*}

\begin{table*}[!t]
\caption{Activity indices and longitudinal magnetic field measurements.} 
\label{tab:indexes}     
\centering                       
\begin{tabular}{l c c c c c c}      
\toprule    
Date & Instrument & $S$-index & $\log R'_\mathrm{HK}$ & H$\alpha$ index & Ca~\textsc{ii} IRT & B$_l$ \\
{[dd-mm-yyyy]} & & & & & & [G]\\
\midrule
12-09-2022 & ESPaDOnS & $0.433\pm0.041$ & $-4.345\pm0.055$ & $0.341\pm0.002$ & $0.899\pm0.002$ & $3.2\pm0.9$\\
22-12-2022 & HARPSpol & \ldots & \ldots & $0.339\pm0.002$ & $\ldots$ & $6.4\pm2.0$\\
24-12-2022 & HARPSpol & \ldots & \ldots & $0.339\pm0.002$ & $\ldots$ & $\ldots$\\
05-02-2023 & HARPSpol & \ldots & \ldots & $0.331\pm0.003$ & $\ldots$ & $\ldots$\\
06-03-2023 & Neo-Narval & $\ldots$ & $\ldots$ & $0.339\pm0.002$ & $0.880\pm0.003$ & $-5.1\pm1.9$\\
07-04-2023 & Neo-Narval & $\ldots$ & $\ldots$ & $0.337\pm0.002$ & $0.885\pm0.003$ & $-0.6\pm1.9$\\
08-04-2023 & Neo-Narval & $\ldots$ & $\ldots$ & $0.338\pm0.002$ & $0.889\pm0.003$ & $-3.3\pm2.0$\\
16-04-2023 & Neo-Narval & $\ldots$ & $\ldots$ & $0.340\pm0.003$ & $0.898\pm0.005$ & $3.6\pm1.9$\\
02-05-2023 & Neo-Narval & $\ldots$ & $\ldots$ & $0.336\pm0.006$ & $0.882\pm0.006$ & $-0.2\pm2.8$\\
05-05-2023 & Neo-Narval & $\ldots$ & $\ldots$ & $0.342\pm0.007$ & $0.910\pm0.006$ & $-9.4\pm2.5$\\
\bottomrule                                 
\end{tabular}
\tablefoot{The error bar for $S$-index and $\log R'_\mathrm{HK}$ is the standard deviation of the measurements from the four sub-exposures. For the H$\alpha$ index, the Ca~\textsc{ii} IRT index and longitudinal magnetic field, they are obtained from formal propagation.}
\end{table*}

\subsection{Longitudinal magnetic field}

We computed the disk-integrated, line-of-sight-projected component of the large-scale magnetic field following \citet{Donati1997} for the snapshot ESPaDOnS observation, the first HARPSpol observation, and the six Neo-Narval observations. We used the general formula as in \citet{Cotton2019}, 
\begin{equation}
\mathrm{B}_l = \frac{h}{\mu_B \lambda_0 \mathrm{g}_{\mathrm{eff}}}\frac{\int vV(v)dv}{\int(I_c-I(v))dv} \,,
\label{eq:Bl}
\end{equation}
where $\lambda_0$ and $\mathrm{g}_\mathrm{eff}$ are the normalisation wavelength and Land\'e factor of the LSD profiles, $I_c$ is the continuum level, $v$ is the radial velocity associated to a point in the spectral line profile in the star's rest frame, $h$ is the Planck's constant, and $\mu_B$ is the Bohr magneton. To express it as in \citet{Rees1979} and \citet{Donati1997}, it is possible to use $hc/\mu_B=0.0214$\,Tm, where $c$ is the speed of light in m\,s$^{-1}$.

The computation was carried out within $\pm20$\,km\,s$^{-1}$ from line centre at $-15.9$\,km\,s$^{-1}$ for both Stokes~$I$ and $V$. We found a value of $3.2\pm0.9$\,G for the ESPaDOnS snapshot observation, $6.4\pm2.0$\,G for the HARPSpol observation and a range between $-9.4$ and $3.6$\,G for the Neo-Narval observations. The median formal error bar is 2.0\,G. In comparison, the value reported in 2010 by \citet{Marsden2014} is $2.5\pm0.5$\,G.
We illustrate the time series of B$_l$ measurements in Fig.~\ref{fig:proxies}. The fast-changing sign is symbolic of a complex magnetic field topology, which is expected for stars of similar properties to HD\,63433 \citep[e.g.][]{Morgenthaler2012,Strassmeier2023}.

\begin{figure}[!t]
    \centering
    \includegraphics[width=0.7\columnwidth]{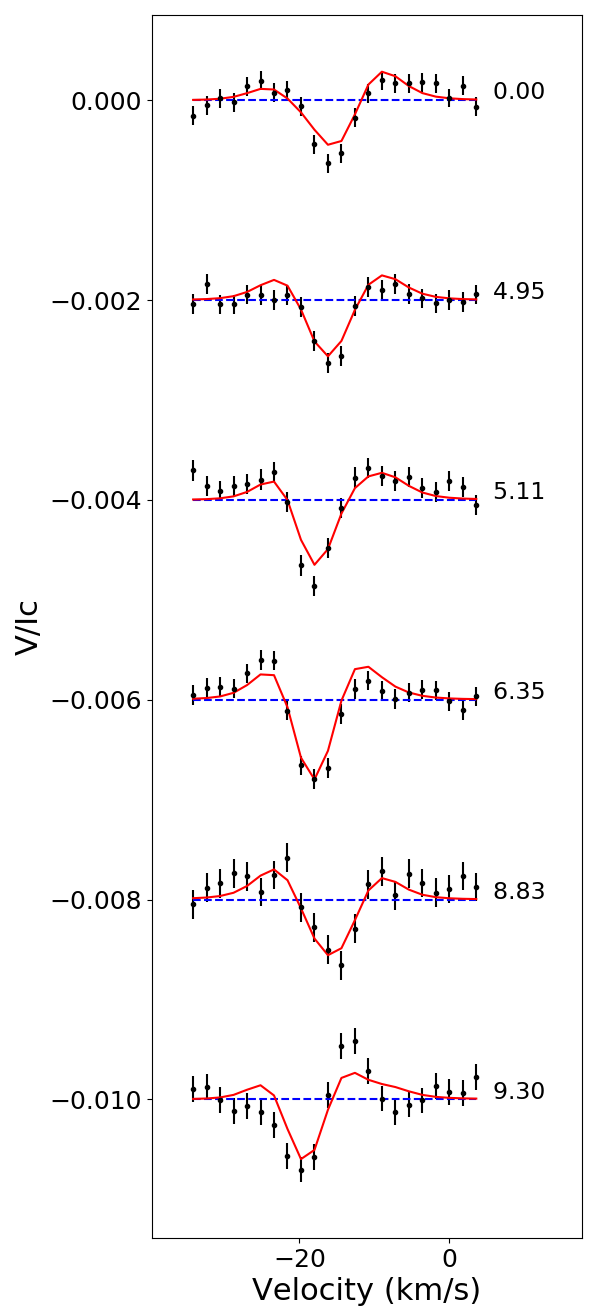}
    \caption{Time series of circularly polarised Stokes profiles obtained with Neo-Narval. Observations are shown as black dots and ZDI models as red lines and they are offset vertically for better visualisation. The number on the right is the rotational cycle (see Eq.~\ref{eq:ephemeris}).}
    \label{fig:stokes_zdi}%
\end{figure}

\subsection{Correlations of activity proxies}

We inspected the correlations between $|$B$_l|$, the H$\alpha$ index and the Ca~\textsc{ii} IRT index for the six Neo-Narval observations. The results are given in Fig.~\ref{fig:proxy_corr}. In general, we caution that these correlations are found with a statistically low number of observations. Therefore, our main aim is to check that we see the expected trends, even for a low number of observations.

The activity indices are computed from unpolarised spectra, which are sensitive to both small and large spatial scales and, thus, to most magnetic features. Instead, the longitudinal field is derived from circularly polarised spectra, which are known to be sensitive to large-scales only, owing to polarity cancellation effects. For this reason, moderate correlations or more complex behaviours between these quantities have been reported for cool dwarfs \citep{Petit2008,Morgenthaler2012,Marsden2014,BoroSaikia2016,Brown2022}.

Here, we obtained a strong positive correlation between H$\alpha$ index and $|$B$_l|$ of Pearson correlation coefficient $\rho=0.95$. The ESPaDOnS and HARPSpol values are not included in the computation of $\rho$, because the corresponding observations were collected months before (see Table~\ref{tab:indexes}), hence intrinsic variability may have occurred (the inclusion of these two points yields $\rho=0.79$, as expected from a loss of coherency of activity over long timescales). For the Ca~\textsc{ii} IRT index versus $|$B$_l|$ case and the H$\alpha$ versus Ca~\textsc{ii} IRT index case, we obtained $\rho=0.75$ and 0.77, respectively. While still strong, the lower level of correlation is likely due to the different chromospheric regions that H$\alpha$ index and Ca~\textsc{ii} IRT index probe \citep[see e.g.][]{Hintz2019}. Figure~\ref{fig:proxy_corr} also demonstrates the high variability of magnetic activity for this star, since the last two measurements (yellow circles beyond rotational cycle 8.0) are at the extremes of the measurements range of the indices (see also Table~\ref{tab:indexes}). 

\section{Zeeman-Doppler imaging}\label{sec:zdi}

We applied ZDI on the six Neo-Narval observations to map the large-scale magnetic field at the surface of HD\,63433. Formally, the field is expressed as the sum of a poloidal and toroidal component, and both components are described by means of spherical harmonic decomposition \citep{Donati2006, Lehmann2022}. The ZDI algorithm generates a time series of Stokes~$V$ profiles and adjusts them to the observations iteratively until a target $\chi^2_r$ is reached. The goal is to fit the spherical harmonics coefficients $\alpha_{\ell,m}$, $\beta_{\ell,m}$, and $\gamma_{\ell,m}$ (with $\ell$ and $m$ being the degree and order of the mode, respectively). The algorithm adopts a maximum-entropy regularisation scheme to reconstruct the magnetic field configuration compatible with the data and with the lowest information content \citep[for more information see][]{Skilling1984,Donati1992,Donati1997}. In practice, we employed the python \texttt{zdipy} code described in \citet{Folsom2018} \footnote{Available at \href{https://github.com/folsomcp/ZDIpy}{https://github.com/folsomcp/ZDIpy}}, assuming weak-field approximation \citep[see e.g.][]{Landi1992}.

We adopted the following stellar input parameters \citep{Mann2020}: inclination of 70$^{\circ}$, projected equatorial velocity ($v_\mathrm{eq} \sin i$) of 7.3\,km\,s$^{-1}$, rotation period of 6.45\,d, and solid body rotation. The value of $v_\mathrm{eq}\sin(i)$ is consistent with other dedicated studies \citep{Rainer2023}. We set a linear limb darkening coefficient of 0.7 \citep{Claret2011} and the maximum degree of harmonic expansion $\ell_\mathrm{max}=5$, consistently with the spatial resolution allowed by $v_\mathrm{eq}\sin(i)$. Using larger values of $\ell_\mathrm{max}$ did not improve the fit since the magnetic energy is stored only in the modes up to $\ell=5$.

The time series of Neo-Narval Stokes~$V$ profiles is shown in Fig.~\ref{fig:stokes_zdi}. The shape of the LSD profiles suggests that the field topology includes a significant contribution from the toroidal component. The last observation at cycle 9.30 seems to have a more antisymmetric shape, which is typical of a more dipolar configuration instead \citep{Morin2008a,Morin2008,Bellotti2023a,Bellotti2023b}. Such variation in profile shape is likely due to stellar rotation rather than intrinsic evolution, considering the short phase separation with the previous observations. As a result, we expect a magnetic feature of negative polarity in the radial field at that phase.

\begin{figure}[!t]
    \centering
    \includegraphics[width=\columnwidth]{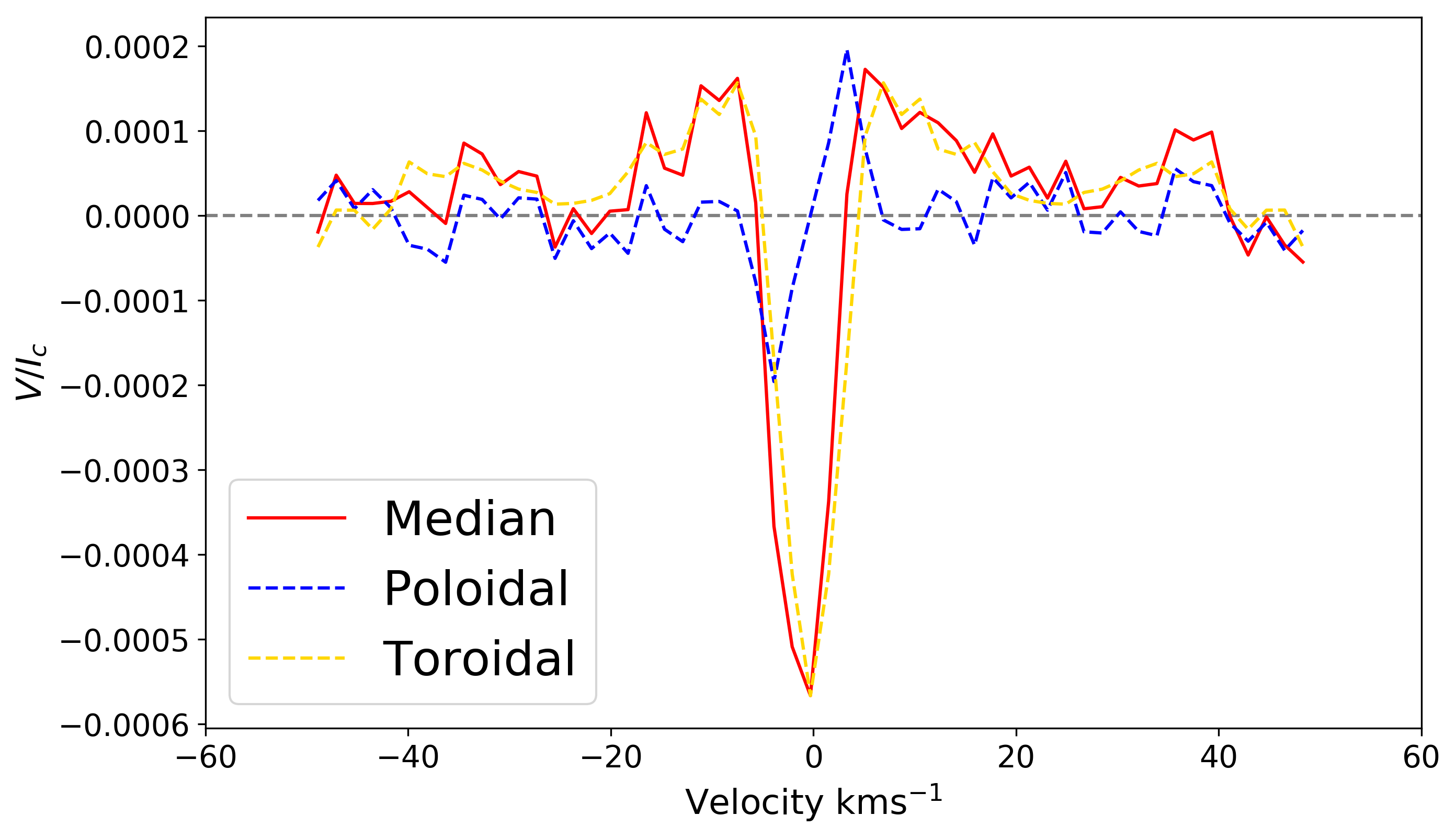}
    \caption{Decomposition of the median Stokes~$V$ LSD profile into its symmetric and antisymmetric components. We note that the median Stokes~$V$ (red line) has mostly a symmetric shape (yellow line) which indicates a strong toroidal component. At the same time, there is still a significant contribution from the antisymmetric shape (blue line), namely, the poloidal component.}
    \label{fig:decomposition}%
\end{figure}

A qualitative assessment of the axisymmetric component of the magnetic field can be obtained by decomposing the median Stokes~$V$ profile into its symmetric and asymmetric components \citep{Lehmann2022}. We show this exercise in Fig.~\ref{fig:decomposition}, where we see that the median Stokes~$V$ shape is mostly symmetric, suggesting a significant contribution from the toroidal component of the field. In parallel, the antisymmetric shape exhibits a strong signal, meaning that the poloidal component is also contributing substantially to the profile shape. From this, one can expect the axisymmetric component of the field to be mostly toroidal \citep[in agreement with][]{Petit2008}, but with  a substantial poloidal contribution as well. This confirms the capability of the symmetric-antisymmetric decomposition to qualitatively determine the main magnetic field component. This could be particularly useful when a large number of stars have to be analysed before a detailed ZDI analysis is carried out. 

\begin{table*}[!t]
\caption{Properties of the magnetic map.} 
\label{tab:zdi_output}     
\centering                       
\begin{tabular}{c c c c c c c c c c}      
\toprule    
B$_\mathrm{mean}$   & B$_\mathrm{max}$ & E$_\mathrm{pol}$/E$_\mathrm{tot}$ & E$_\mathrm{tor}$/E$_\mathrm{tot}$  & E$_\mathrm{dip}$/E$_\mathrm{pol}$   & E$_\mathrm{quad}$/E$_\mathrm{pol}$  & E$_\mathrm{oct}$/E$_\mathrm{pol}$   & E$_\mathrm{axi}$/E$_\mathrm{tot}$ & E$_\mathrm{axi,pol}$/E$_\mathrm{pol}$ & E$_\mathrm{axi,tor}$/E$_\mathrm{tor}$ \\
{[G]} & [G] & [\%] & [\%] & [\%] & [\%] & [\%] & [\%] & [\%] & [\%]\\
\midrule
23.8$^{+6.2}_{-4.4}$ & 54.4$^{+19.1}_{-15.3}$& 46.3$^{+15.3}_{-9.4}$& 53.7$^{+9.4}_{-15.3}$& 30.2$^{+25.2}_{-4.7}$& 25.6$^{+8.9}_{-3.8}$ & 15.2$^{+0.7}_{-10.6}$& 65.7$^{+24.6}_{-17.3}$& 37.7$^{+39.1}_{-14.8}$& 89.9$^{+8.3}_{-5.9}$\\
\bottomrule                                 
\end{tabular}
\tablefoot{The following quantities are listed: mean absolute magnetic strength, maximum value of the magnetic field modulus, poloidal and toroidal magnetic energy as a fraction of the total energy, dipolar, quadrupolar, and octupolar magnetic energy as a fraction of the poloidal energy, axisymmetric magnetic energy as a fraction of the total energy, poloidal axisymmetric energy as a fraction of the poloidal energy, and toroidal axisymmetric energy as a fraction of the toroidal energy. The variation bars were computed by reconstructing ZDI maps and including the uncertainties on the input stellar parameters.}
\end{table*}

\begin{figure*}[!t]
    \centering
    \includegraphics[width=\textwidth]{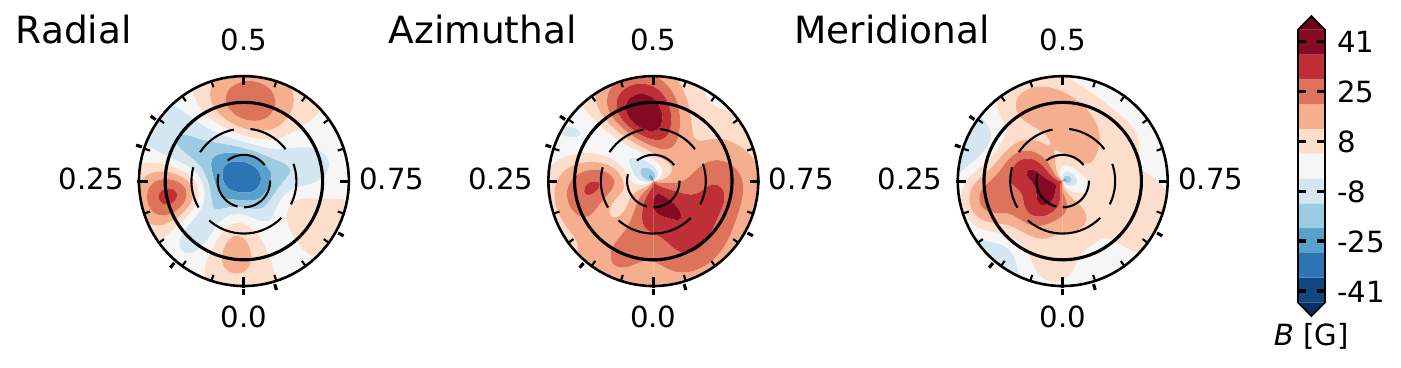}
    \caption{ZDI reconstruction in flattened polar view of the large-scale field of HD\,63433. From the left, the radial, azimuthal, and meridional components of the magnetic field vector are illustrated. The radial ticks are located at the rotational phases when the observations were collected, while the concentric circles represent different stellar latitudes: -30\,$^{\circ}$, +30\,$^{\circ}$, and +60\,$^{\circ}$ (dashed lines), as well as the equator (solid line). The geometry is complex, with a significant toroidal fraction and the poloidal component distributed mainly in the dipolar, quadrupolar, and octupolar modes. The colour bar encapsulates the magnetic field strength, up to a maximum of 55\,G.}
    \label{fig:zdi}%
\end{figure*}

The Stokes~$V$ profiles were fit down to $\chi^2_r=1.2$, from an initial value of 6.5 (see Fig.~\ref{fig:stokes_zdi}). The value was chosen to prevent artificial injection of magnetic energy in high order modes or reconstruction of substantially weak fields, which would be symbolic of overfitting and underfitting of the Stokes~$V$ profiles, respectively \citep[see e.g.][]{Alvarado-Gomez2015}. Lowering the $\chi^2_r$ further would likely require the inclusion of differential rotation as intrinsic variability mechanism, but with our sparse observations it cannot be constrained. 

The properties of the magnetic map are given in Table~\ref{tab:zdi_output} and the map is shown in Fig.~\ref{fig:zdi}. The mean magnetic field strength is B$_\mathrm{mean}=24$\,G, with the toroidal component accounting for 54\% of the magnetic energy and the poloidal component for 46\%. The dipolar mode accounts for 30\% of the poloidal energy and both the quadrupolar and quadrupolar modes store a significant fraction, namely, 26\% and 15\%.

To set the uncertainties on the magnetic field properties, we estimated variation bars following \citet{Mengel2016}, \citet{Fares2017}, and \citet{Bellotti2023a}. The idea is to vary the input parameters, namely inclination, $v_\mathrm{eq}\sin(i)$, and rotation period to plausible values consistent with the literature. One at a time, we set these input parameters to the extremes of the range of values reported in the literature, and reconstructed the corresponding map of the magnetic field. The variation bars on the field characteristics are then the maximum deviations of these new maps from the reconstruction with the optimised parameters. These variation bars do not capture the actual error bars, but, rather, the sensitivity of the reconstructed properties to input parameters.

Quantitatively, \citet{Mann2020} constrained the stellar inclination to be larger than $74^{\circ}$; hence, we reconstructed maps with $i=60^{\circ}$ and $i=80^{\circ}$. For $v_\mathrm{eq} \sin i$, we used 6.5\,km\,s$^{-1}$ and 7.9\,km\,s$^{-1}$ \citep{Damasso2023} and for P$_\mathrm{rot}$ we used 6.0\,d \citep{Mallorquin2023} and 7.6\,d \citep{Marsden2014}. We adopted a target $\chi^2_r$ of 1.2 in every case, except for P$_\mathrm{rot}=6.0$\,d for which we used $\chi^2_r=2.4$. This already indicates that using such value of rotation period degrades significantly the magnetic field model. The variation bars of the magnetic characteristics are driven predominantly by the uncertainties on the rotation period, since the variations in inclination and $v_\mathrm{eq} \sin i$ have led to magnetic maps that are reasonably consistent with the reconstruction when using the optimised parameters.

To check the robustness of our results, we reconstructed a map excluding the last observation. The only change is in the energy fraction in the poloidal component, as it decreases to 25\%. This is not surprising considering that the shape of Stokes~$V$ in the last spectropolarimetric observation is the most similar to a typical poloidal-dipolar field \citep[e.g.][]{Morin2008,Bellotti2023a}. The average field strength changes only by 1\,G, while the dipolar, quadrupolar and octupolar energy fraction by less than 2\% and the axisymmetric fraction by less than 15\%. Such changes are encompassed by the variation bars we estimated in this work.

\section{ZDI-driven stellar wind models}\label{sec:wind}

We created the stellar wind models using the Alfvén wave solar model~\citep[AWSoM,\ ][]{2013ApJ...764...23S,2014ApJ...782...81V}, which is a component in the Space Weather Modelling Framework \citep[SWMF,\ ][]{2005JGRA..11012226T,2012JCoPh.231..870T}. This model solves the ideal two-temperature magnetohydrodynamic equations, supplemented by a pair of equations describing the propagation of Alfvén wave energy along magnetic field lines. The model extends from the chromosphere, where the temperature is \qty{5e4}{\kelvin} and the density is \qty{2e17}{\per\cubic\meter}, through the transition region to the stellar corona. The model is driven by the energy flux of Alfvén waves, which crosses the inner boundary at a rate proportional to the local magnetic field strength at the stellar surface. The Poynting flux-to-field ratio is set to $1.1\times10^6$~W~m$^{-2}$~T$^{-1}$ and the turbulence correlation length is $1.5\times10^5$~m~T$^{1/2}$. The radial component of the surface magnetic field is held to the \(B_r\) values in Fig.~\ref{fig:zdi}, while the transverse components are left to evolve with the numerical solution. The AWSoM model was configured using solar values that have been shown to reproduce solar conditions~\citep{2015MNRAS.454.3697M,2019ApJ...872L..18V,2019ApJ...887...83S}. The coronal heating due to Alfvén waves is likely to resemble that of the Sun, according to the non-thermal velocity measurements of sun-like stars of similar rotation periods in \citet{BoroSaikia2023}. For a more detailed description of the methodology behind the wind models, we refer to ~\citet{2021MNRAS.506.2309E,2022MNRAS.510.5226E,2023MNRAS.524.2042E}.

\begin{figure}
    \includegraphics[width=\columnwidth]{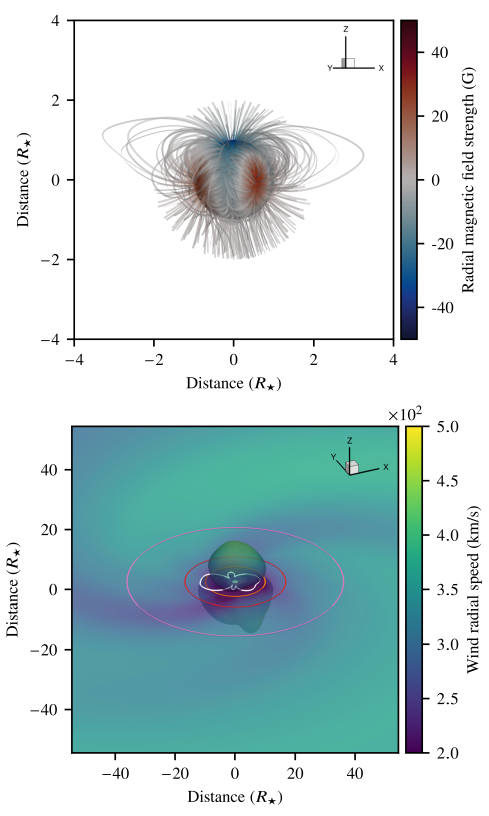}
    \caption{Simulated environment of HD\,63433. {\it Top}: Steady-state coronal magnetic field structure obtained from our ZDI-driven stellar wind model. Open magnetic field lines are truncated at two stellar radii, while the closed field lines are not truncated. The star is seen equator on at the rotational phase 0.35. The large-scale structure exhibits a mostly dipolar topology. The colours are the same as Fig.~\ref{fig:zdi}. {\it Bottom}: Alfvén surface and wind radial speed. The two-lobed Alfvén surface is characteristic for a dipole-dominated coronal magnetic field. The intersection of the Alfvén surface and the stellar equatorial plane is shown as a white curve, while the the planetary orbits are shown in orange, red, and pink. The orbital plane is colour-coded by the wind radial speed.}
    \label{fig:wind-alfven}
\end{figure}

The top panel of Fig.~\ref{fig:wind-alfven} shows the stellar magnetic field embeded in the stellar wind equator-on. The large scale coronal magnetic field resembles a dipole with closed regions pushed towards the positive $z$ axis. We also see arch-like structures where the magnetic field lines connect regions of opposite polarity, which are the result of quadrupolar and octupolar components of the magnetic field as may be seen also in the radial field plot of Fig.~\ref{fig:zdi}. In our model, the wind mass loss rate is $1.1\times10^{-13}\,\text{M}_\odot/\text{yr}$, which is about five times the solar wind mass loss rate.

The bottom panel of Fig~\ref{fig:wind-alfven} shows the Alfvén surface, where the stellar wind speed equals the Alfvén wave velocity. This quantity is crucial to understand the regime in which magnetic or inertial forces dominate, within or outside the surface, respectively \citep[e.g.][]{Vidotto2021}. A practical quantity in this context is the Alfvén Mach number, which is defined as the ratio of the wind speed, \(v_\textsc{sw}\), to the Alfvén wave speed of \(v_\textsc{a} = B/\sqrt{4\pi \rho_\textsc{sw}}\), where \(B\) is the local magnetic field strength and \(\rho_\textsc{sw}\) is the local wind density. The Alfvén Mach number quantifies whether the planetary motion is super-(M$_\mathrm{A}>1$) or sub-Alfvénic (M$_\mathrm{A}<1$). The Alfvén surface corresponds to M$_\mathrm{A}=1$. 

The orbits of planet d, b, and c are shown in the bottom panel of Fig.~\ref{fig:wind-alfven} as concentric curves at 9.9, 16.8, and 36.1\,R$_\star$ \citep{Damasso2023,Mallorquin2023,Capistrant2024}. For the three planets, we assume that the planets have circular orbits in the equatorial plane of the star, consistently with the literature estimates within error bars \citep{Mann2020,Damasso2023,Capistrant2024}. The orbit of the innermost planet in our model crosses briefly the Alfvén surface into the sub-Alfvénic regime; we return to this in Fig.~\ref{fig:alfven-number} and  discuss it further in Sect.~\ref{sec:discussion}. 

\begin{figure}
    \includegraphics{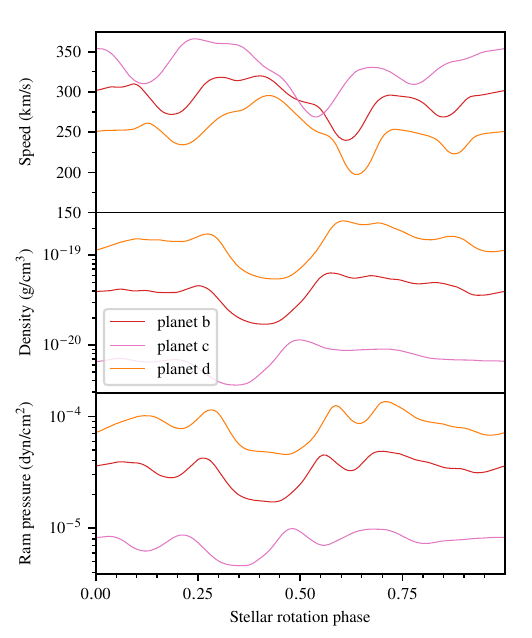}
    \caption{
        Stellar wind conditions in the planetary frame. From the top: Wind speed, density, and ram pressure are shown as a function of stellar rotation phase. Planets b, c, and d are colour-coded in red, pink, and orange.
    }\label{fig:phase-triple}
\end{figure}

\begin{figure}
    \includegraphics{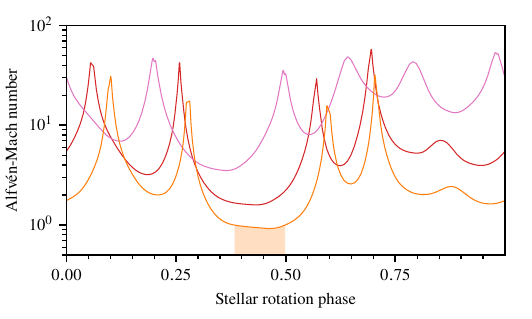}
    \caption{
        Alfvén Mach number computed in the planetary frame as a function of stellar rotation phase. The colours and labels are the same as in Fig.~\ref{fig:phase-triple}. The phase interval where our model's planet d crosses into the sub-Alfvénic regime is shown as a green shaded region.
    }\label{fig:alfven-number}
\end{figure}

Figure~\ref{fig:phase-triple} shows the wind speed \(v_\textsc{sw}\), density, \(\rho_\textsc{sw}\), and wind ram pressure, \(\rho_\textsc{sw} v_\textsc{sw}^2\), in the planetary frame, as a function of stellar rotation phase. We note the presence of three dips in the wind speed curves. These are correlated to the three arms of the wind spiral shape which, in turn, are due to stellar rotation (see Fig.~\ref{fig:wind-alfven}). Such spiral features were already reported, and tend to build more arms if the field complexity is larger \citep{2021MNRAS.506.2309E}. The wind density experienced by each planet can vary by a factor of $\sim5$ along planetary orbits. For the ram pressure, the variation is somewhat smaller than that of the density. 

Figure~\ref{fig:alfven-number} shows the Alfvén Mach number (M$_\mathrm{A}$) computed in the planetary frame as a function of stellar rotational phase. In our model, we see that planet d experiences sub-Alfvénic winds at stellar rotation phases 0.38--0.50, whereas planets b and c orbit in super-Alfvénic motions. In the next section, we discuss what these locations translate to in terms of star-planet interactions and their observable signatures. 

\section{Summary and discussion}\label{sec:discussion}

We have characterised the magnetic field and environment of the active, young G5 dwarf star HD\,63433, which is known to host two sub-Neptunes and an Earth-sized planet. We computed magnetic activity diagnostics such as the chromospheric $S$ index, $\log R'_\mathrm{HK}$, H$\alpha$, Ca~\textsc{ii} infrared triplet, and longitudinal magnetic field, using observations from ESPaDOnS, HARPSpol, and Neo-Narval. We then reconstructed the large-scale magnetic field by means of ZDI and used it as boundary condition for the simulation of the system's stellar wind from the corona to the orbital distances of the planets b, c, and d, following the method presented in \citet{2021MNRAS.506.2309E}. In general, the observations and analyses were conducted as part of a survey dedicated to the spectropolarimetric characterisation of bright stars in the current list of potential $Ariel$ targets, in order to efficiently construct the final target list before the launch and, consequently, optimize the science outcome of the mission.

\begin{figure*}[!ht]
    \centering
    \includegraphics[width=\textwidth]{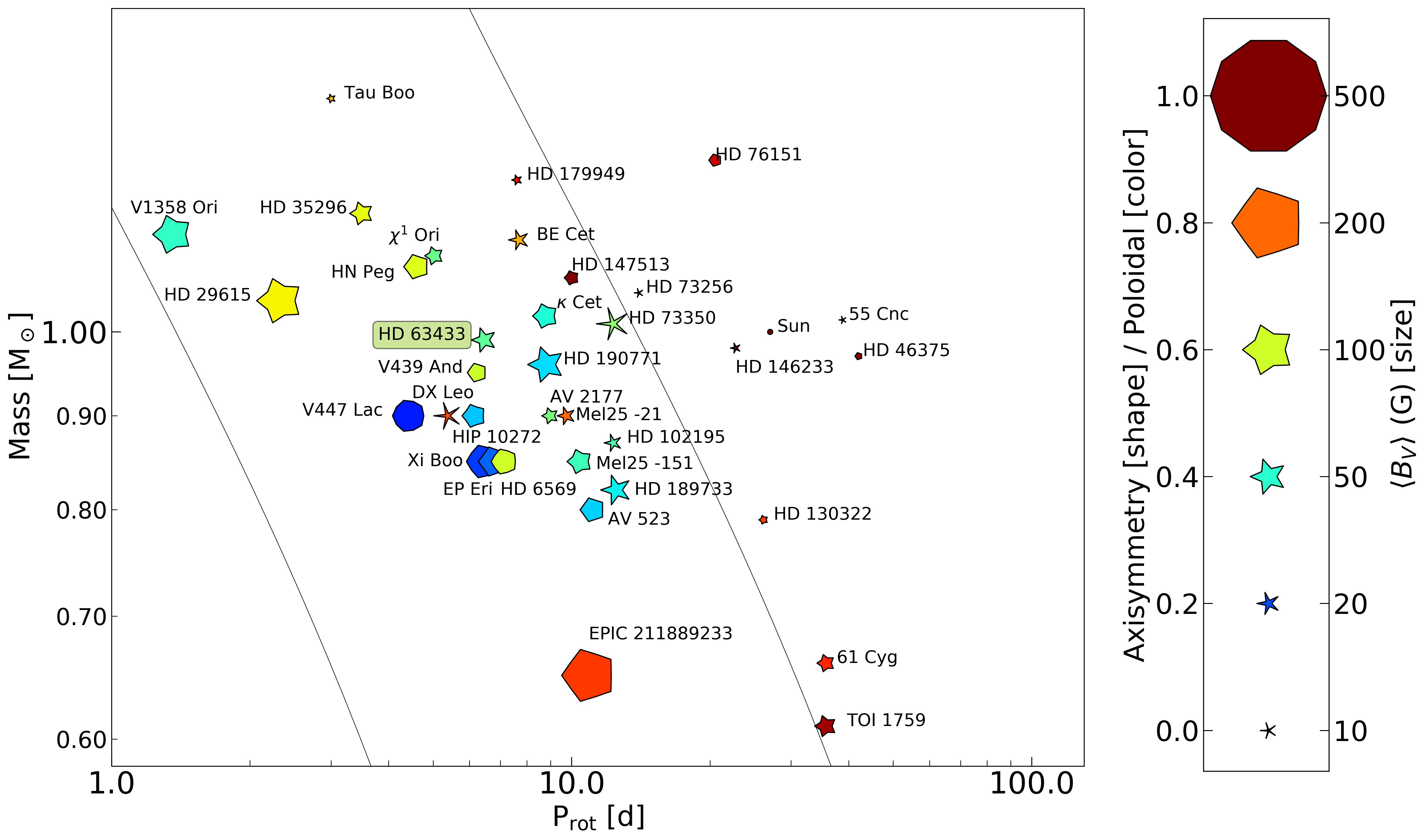}
    \caption{Properties of the magnetic topologies for cool, main sequence stars obtained via ZDI. The location of HD\,63433 is highlighted. The $y$- and $x$-axes represent the mass and rotation period of the star, and iso-Rossby number curves are overplotted using the empirical relations of \citet{Wright2018}. 
    The symbol size, colour and shape encodes the ZDI average field strength, the poloidal and toroidal energy fraction, and axisymmetry. Data entering the plot are taken from \citealt{Petit2008,Morgenthaler2012,Fares2013,BoroSaikia2015,DoNascimento2016,BoroSaikia2016,Hussain2016,Alvarado-Gomez2015,Folsom2016,Folsom2018,Folsom2020,Klein2021,Willamo2022,Martioli2022}.}
    \label{fig:megaconfusogram}%
\end{figure*}

We estimated $S=0.433\pm0.041$ and $\log R'_\mathrm{HK}=-4.345\pm0.055$, which are compatible with the measurements of other Sun-like stars with analogous spectral type, mass, rotation period, and age ($\xi$\,Boo\,A, \citealt{Morgenthaler2012,Strassmeier2023}; HD\,1237, \citealt{Alvarado-Gomez2015}; HD\,147513, \citealt{Hussain2016}; $\kappa$\,Cet, \citealt{DoNascimento2016}). In comparison, the average values for the Sun at cycle maxima and minima are $\log R'_\mathrm{HK}=-4.905$ and $-4.984$ \citep{Egeland2017}. 

The longitudinal magnetic field of HD\,63433 varies between $-11$ and 5\,G, which is typical for solar-like stars. For instance, \citet{Morgenthaler2012} found variations between $-2.3$ and 22\,G for $\xi$\,Boo\,A, \citet{BoroSaikia2016} reported values spanning $-20$ and 10\,G across different epochs of HN~Peg, and \citet{Alvarado-Gomez2015} reported a range between $-10$ and 10\,G for HD\,1237. The large-scale field is three times stronger (in absolute value) than the Sun, whose large-scale field values can reach 3-4\,G \citep{Kotov1998}.

The reconstructed magnetic field topology of HD\,63433 is complex, with a significant amount of energy stored in the toroidal and poloidal components (see Sect.~\ref{sec:mag_characterisation}). The average field strength in absolute value is around 23\,G. This is of the same order of magnitude as estimated by \citet{Zhang2022} who employed the age-magnetic field strength law of \citet{Vidotto2014} to include the effects of stellar magnetic fields in the outflow models of planet b and c. In our model, the derived stellar wind mass loss rate is $1.1\times10^{-13}\,\text{M}_\odot/\text{yr}$, which is 2.9 times smaller than the value derived  from the simulations of \citet{Zhang2022}. The stellar wind can affect atmospheric escape in exoplanets in several ways, from potentially reducing escape rates \citep{Vidotto2020} and altering the signatures of atmospheric evaporation through spectroscopic transits \citep{Carolan2021}, even to possibly creating a tenuous atmosphere through sputtering processes \citep{Vidotto2018}. Recently, \citet{Zhang2022} modelled the escaping atmospheres of HD\,63433\,b and c, including the effects of the interaction with the stellar wind. So far, no model exists for the escaping atmosphere of the recently discovered innermost planet HD\,63433\,d. With the optimally constrained stellar wind properties presented in our work, future atmospheric models of HD\,63433\,d will be able to better pin-point the effects of the wind of the host star on the upper atmosphere of HD\,63433\,d, as well as potential signatures in Ly-$\alpha$ and He~\textsc{i} transits. Such models are important to assess the presence and evolution of the atmosphere of HD\,63433\,d \citep{Kubyshkina2022}.

Finally, we illustrate the properties of the field geometry of HD\,63433 compared to other stars in Fig.~\ref{fig:megaconfusogram}. Although the number of observations used for tomographic inversion is limited, the magnetic field exhibits similarities with other Sun-like stars of similar properties and remarkably with the field map of $\xi$\,Boo\,A from 2008 \citep{Morgenthaler2012} because it also exhibits a dominating polarity at the north pole, as well as lower latitude magnetic spots.

In terms of evolution, \citet{Maldonado2022} analysed activity indices and photometric light curves of HD\,63433, and estimated the lifetime of active regions to be approximately 60\,d. Additionally, \citet{Lehtinen2016} reported cyclic photometric variability on a 2.7\,yr timescale and a less significant variation of approximately 8.0\,yr. If we maintain the analogy with $\xi$\,Boo\,A, \citet{Morgenthaler2012} reported fast variations over timescales of six months of the field topology and complexity, from a poloidal-dominated to a toroidal-dominated regime. Overall, this shows that HD\,63433 could manifest evident and important variability, which motivates additional spectropolarimetric monitoring. 

From the stellar environment simulations, we obtained a largely dipolar coronal magnetic field, aligned with the stellar axis of rotation. We also see arch-like structures where the magnetic field lines connect regions of opposite polarity, resulting from the quadrupolar and octupolar components of the magnetic field. Using our wind model, we computed the location of the Alfvén surface where the wind speed equals the Alfvén wave velocity. The Alfvén surface is two-lobed, which is characteristic for a dipole-dominated coronal magnetic field. 

We found that the outermost two planets orbit in the super-Alfvénic region. In this regime, a bow shock between the stellar wind and a planetary magnetosphere (if present) can be generated and the stellar wind is deflected around the planet~\citep{1931TeMAE..36...77C,2009ApJ...703.1734V, Vidotto2010}. Also, the formation of a tail-like structure of evaporated planetary material can occur \citep[e.g.][]{Schneiter2007,Villareal2018,Zhang2022}. Star-planet interactions in super-Alfvénic regime are expected to produce auroral radio emission from the planetary magnetosphere, similarly   to Solar System planets \citep{Zarka2001}.

The innermost Earth-sized planet is located in the trans-Alfvénic region, as our modelling indicates that the stellar wind conditions along its orbit may switch between sub- and super-Alfvénic. Within the sub-Alfvénic regime, the planet can directly link to the stellar magnetic field, and its motion can perturb the field lines producing Alfvén waves travelling starward; alternatively, it can induce magnetic reconnection events (if magnetised) or a combination of both \citep[e.g.][]{Neubauer1998,Ip2004,Strugarek2015,Kavanagh2022}. In turn, these phenomena generate detectable signatures such as coronal radio emission on the star \citep{Saur2013,Kavanagh2021} and local chromospheric flux enhancement \citep{Shkolnik2008,Lanza2009}.

We note that the magnetic field strength recovered from ZDI could be underestimated \citep{Lehmann2019}. On one side, the latitudinal coverage of the star is limited by definition, as we can only model the disk intersected by the line of sight. On the other side, the low number of observations combined with the poor longitudinal coverage of the stellar rotation may prevent us from reconstructing additional large-scale magnetic features. For these reasons, a more robust wind model will be provided in the future with an updated map built from a denser spectropolarimetric monitoring. Although spectropolarimetric time series with a scarce sampling may limit the reconstruction of the large-scale magnetic topology, the first-order information that we can derive from them is still relevant feedback for $Ariel$ target planning. They allow us to assess the magnetic activity level of the star and its variation over distinct timescales: from rotational modulation to long-term evolution, potentially in the form of a magnetic cycle, which could be seen as a magnetic polarity reversal even from multiple epochs in which the rotational phase coverage is not optimal \citep[see Fig.~1][for example]{BoroSaikia2016}.

This work has been developed within the science development of the ESA $Ariel$ space mission, for which more magnetic activity characterisation studies will be presented in the upcoming future. It represents the start of a campaign dedicated to the magnetic characterisation of stars in the current list of $Ariel$ targets \citep{Edwards2022}. In this regard, additional observations of HD\,63433 are necessary to understand the long-term evolution of its large scale magnetic field. This will be also valuable to our understanding of the observational signatures of star-planet interactions due to the correlated long-term evolution of the Alfven surface. Indeed, it would be interesting to observe the star along its likely magnetic cycle to characterize how the Alfvén surface moves with respect to the planetary orbits. Observations in the UV range will also be essential, as X-ray observations show a difference of around two orders of magnitude between the Sun and HD\,63433. This raises the question of how the UV wavelength range differs and the consequences on the atmospheric photochemistry, heating, and loss. Ultimately, this information can provide practical constraints for the observing strategies of $Ariel$, for such targets as the HD\,63433 system. Once the mission is launched, the goal of our spectropolarimetric campaign will become more of a follow-up work, aimed at complementing the observations already performed and providing new insights for the interpretation of $Ariel$ data.

\begin{acknowledgements}

This publication is part of the project "Exo-space weather and contemporaneous signatures of star-planet interactions" (with project number OCENW.M.22.215 of the research programme "Open Competition Domain Science- M"), which is financed by the Dutch Research Council (NWO). This work used the Dutch national e-infrastructure with the support of the SURF Cooperative using grant nos. EINF-2218 and EINF-5173. SB acknowledges funding from the SCI-S department of the European Space Agency (ESA), under the Science Faculty Research fund E/0429-03. AAV and DE acknowledge funding from the European Research Council (ERC) under the European Union's Horizon 2020 research and innovation programme (grant agreement No 817540, ASTROFLOW). SBS was supported by the Austrian Science Fund (FWF) Lise Meitner grant M2829-N. CD acknowledges financial support from the INAF initiative ``IAF Astronomy Fellowships in Italy'', grant name \textit{GExoLife}. GM acknowledges the support of the ASI- INAF agreement 2021-5-HH.0 and  PRIN MUR 2022 (project No. 2022J7ZFRA,  EXO-CASH). Based on observations obtained at the Canada-France-Hawaii Telescope (CFHT) which is operated by the National Research Council of Canada, the Institut National des Sciences de l'Univers of the Centre National de la Recherche Scientique of France, and the University of Hawaii. Based on observations collected at the European Southern Observatory under ESO programme 110.24C8.002 (P.I. S. Bellotti). We thank the TBL team for providing service observing with Neo-Narval. This work has made use of the VALD database, operated at Uppsala University, the Institute of Astronomy RAS in Moscow, and the University of Vienna; Astropy, 12 a community-developed core Python package for Astronomy \citep{Astropy2013,Astropy2018}; NumPy \citep{VanderWalt2011}; Matplotlib: Visualization with Python \citep{Hunter2007}; SciPy \citep{Virtanen2020} and PyAstronomy \citep{Czesla2019}.

\end{acknowledgements}

%
%

\bibliographystyle{aa}
\bibliography{biblio}

\end{document}